\documentclass[12pt]{iopart}

\usepackage{bm}        
\usepackage{xcolor}

\usepackage{dcolumn}   
\usepackage{bm}        
\usepackage{multirow}

\usepackage{amsmath}
\usepackage{amsfonts}
\usepackage{amssymb}
\usepackage{makeidx}
\usepackage{graphicx}
\usepackage{anysize}
\usepackage{hyperref}
\usepackage{slashed}
\usepackage{bbm}

\usepackage{xcolor}

\usepackage{ulem} 

\newenvironment{mymathbox}
{\par\smallskip\centering\begin{lrbox}{0}%
\begin{minipage}[c]{0.8\textwidth}}
{\end{minipage}\end{lrbox}%
\framebox[0.9\textwidth]{\usebox{0}}%
\par\medskip
\ignorespacesafterend}
\newcommand{\bb}{\begin{mymathbox}}
\newcommand{\eb}{\end{mymathbox}}

\newcommand{\be}{\begin{equation}}
\newcommand{\ee}{\end{equation}}
\newcommand{\ba}{\begin{eqnarray}}
\newcommand{\ea}{\end{eqnarray}}

\newcommand{\np}{{\bf      p}}

\newcommand{\npsi}{{\bf \npsi}}

\newcommand{\Psib}{\overline{\Psi}}

\newcommand{\bma}{\begin{pmatrix}}
\newcommand{\ema}{\end{pmatrix}}

\begin{document}

\title[Effects of two-body currents in the one-particle one-hole electromagnetic responses]{Effects of two-body currents in the one-particle one-hole electromagnetic responses within a relativistic model}


\author{T. Franco-Munoz$^1$, R.~Gonz\'alez-Jim\'enez$^1$ and J.M.~Ud\'ias$^1$}
\address{$^1$Grupo de F\'isica Nuclear, Departamento de Estructura de la Materia, F\'isica T\'ermica y Electr\'onica, Facultad de Ciencias F\'isicas, Universidad Complutense de Madrid and IPARCOS, CEI Moncloa, Madrid 28040, Spain}
\vspace{10pt}

\begin{indented}
\item[]\today
\end{indented}

\begin{abstract}

Longitudinal ($R_L$) and transverse ($R_T$) responses from inclusive  electron scattering from carbon 12 and calcium 40 nuclei are computed within a fully relativistic and unfactorized model for the initial  and final states, and one- and two-body current operators  leading to the one-particle one-hole responses. 
We find that the two-body contributions have no effect on $R_L$ but they increase $R_T$ by up to 30\%, depending on the energy and momentum transfer.  
 Inclusive cross sections have also been computed. In this case, the increase of $R_T$ due to two-body currents will translate into an increase in the cross-sections depending on the degree of transversity of each kinematic.

The comparison with carbon data is good for the responses and the cross sections. 
In the case of calcium, while the model compares well with the cross section data, the agreement with the responses is generally poor. However, the inconsistencies between different data sets for the separate responses in this nucleus points to uncertainties underlying the procedure to extract the responses that are not considered (or largely underestimated) in the experimental error bars.

Our calculation is fully relativistic  and considers within the full  quantum mechanical description both the initial and final nucleon states involved in the process.  
We also show that it is essential to go beyond the plane-wave approach, since incorporating the distortion of the nucleons while making the initial and final states orthogonal, allows to reproduce both the shape and magnitude of the cross section data and carbon responses.
The good agreement with the electron scattering experimental data supports the use of this approach to describe the analogous neutrino-induced scattering reaction.

\end{abstract}


\section{Introduction}

Electron scattering is one of the most precise and efficient  methods to  determine the internal structure of atomic nuclei. Electron scattering experiments are regaining interest nowadays~\cite{Khachatryan21,Gu21} due to the exponentially growing experimental effort in neutrino-nucleus  scattering, related to neutrino oscillation experiments, and thus the need to understand the nuclear interactions in the target nuclei which otherwise would prevent the extraction of neutrino properties from these experiments~ \cite{Alvarez-Ruso18, Katori17}.  At the very least, the theory employed to describe nuclear effects when analyzing neutrino-nucleus experiments should compare fairly well when put at test against the available electron scattering data under similar kinematics. Of particular interest are the few experimental data available on which not only the cross sections, but also the different contributions from the nuclear responses are separated, as these, while should be similar in electron and neutrino scattering, would be combined differently in both processes. That means, good agreement to electron scattering cross-sections would not be enough to ascertain the reliability of the model, while agreement to every separate nuclear response would be a much more compelling evidence of adequacy of the model.

Describing lepton-nucleus scattering is a complex many-body problem. The energy region of interest for accelerator-based neutrino experiments (e.g. MiniBooNE~\cite{MiniBooNECC13b}, T2K~\cite{T2K20}, MicroBooNE~\cite{MicroBooNE20}, MINERvA~\cite{MINERvA20}, NOvA~\cite{NOvA19}, HyperK~\cite{HyperK15} and DUNE~\cite{DUNE16}) corresponds to incident neutrino energies ranging from a few hundreds of MeVs to tens of GeVs. Among all the mechanisms involved in the energy regime of the above mentioned neutrino experiments, the quasielastic (QE) channel is the dominant  one in T2K  and MicroBooNE, and a major contribution  in MINERvA, NOvA and DUNE. In this work we focus on the QE process, it corresponds to the lepton being scattered by a single nucleon that is consequently ejected from the target nucleus.

The modeling of the scattering process in this region is virtually always performed within the first-order Born approximation, in which one considers that  only one boson is exchanged between the lepton and the nuclear system. Further, most often the impulse approximation is employed in the QE region, meaning that the lepton interacts only with  one nucleon, that is subsequently knocked out from the nucleus. All other nucleons would be spectators contributing just to recoil energy and momentum in the process. It is also important to stress that due to the fact that  the energy ($\omega$)  and momentum (q) transfer between the lepton and the nucleus are in some cases comparable or larger than the mass scale set by the nucleon mass, relativistic effects,  both in the kinematics and dynamics, are relevant.

To a large extent, this process involves a low-energy hole and a high-energy particle, thus we expect pions to play an important role. We extend the treatment of QE scattering, based on a one-body current operator, and include one-pion exchange effects by incorporating a two-body meson-exchange current operator. In this work, meson-exchange currents (MEC) include the dominant Delta-resonance mechanism (Fig.~\ref{fig:delta}, electromagnetic excitation of the $\Delta$(1232) resonance and its subsequent decay into $N\pi$) and the background contributions deduced from the chiral perturbation theory Lagrangian of the pion-nucleon system \cite{Scherer12} (Fig.~\ref{fig:background}, ChPT background or, simply, background terms in what follows). In particular, we have studied the contribution of the two-body meson-exchange currents to the one-particle one-hole (1p-1h) response computed in the impulse approximation. 

Previous works have computed the contribution of pion exchange currents to the 1p-1h and 2p-2h responses within different frameworks
~\cite{VanOrden81,Ryckebusch94,DePace03,Martini09,Nieves11,Amaro20,Amaro94,VanderSluys95,Boffi90-letter,Lovato16,Dekker94,Umino95,Umino95b,Amaro02,Lovato23}. 
There is a consensus that the effect of MEC in the 2p-2h sector leads to a significant contribution in the dip region between the QE and the delta resonance peaks~\cite{VanOrden81,Ryckebusch94,DePace03,Martini09,Nieves11,Amaro20}.
The role of MEC in the 1p-1h responses has been, however, much less explored. In \cite{Amaro94}, within a non-relativistic shell model that  incorporates final-state interactions, it was obtained that  the two-body  current produced a small  decrease  of the transverse response  ($R_T$). 
In  \cite{VanderSluys95}, using a similar nuclear model, it was found that the two-body currents enhance $R_T$ by around 20-30\%. 
In both approaches, by construction, the two-body operator does not affect the longitudinal response ($R_L$). More recently, the {\it ab initio} model of \cite{Lovato16} has confirmed the essential role of two-body mechanisms to describe the electromagnetic responses of light nuclei. 
These previous works employed non-relativistic approaches subjected to hold only at relatively low momentum transfer. Hence, MEC contributions to 1p-1h final states have also been studied within fully relativistic frameworks, for example \cite{Dekker94,Umino95,Umino95b,Amaro02}. However, these approaches, based on the relativistic Fermi gas model, over-simplify the complexity of the nuclear structure and dynamics.  
Other relativistic approach is the one of \cite{Lovato23}, based on the factorized spectral function formalism, they found that the two-body contributions enhance the transverse electromagnetic response functions.
Spectral function approaches incorporate nuclear complexity beyond mean field predictions, thus representing a more realistic response of the nucleus to the external lepton probe. They require, however, to compute the cross section in a factorized fashion, which is not always a good approximation~\cite{Caballero98a,Nikolakopoulos19} and that precludes including final state interactions in a consistent way~\cite{Gonzalez-Jimenez19}. However, a representation of the spectral function can be made without resorting to factorization \cite{Gonzalez-Jimenez22}, which will be exploited in this work to compute the inclusive cross sections and responses.

The results presented in this work are computed within a fully relativistic and quantum mechanical framework, where the initial state is described by an unfactorized representation of the spectral function based on single-particle solutions of relativistic mean-field (RMF) model \cite{Serot86}. The final state is described as a solution of the Dirac equation for the final nucleon in the presence of relativistic potentials. This way we obtain a realistic description of the scattering process that can be applied in the entire kinematical region of interest for electron and neutrino scattering.  
Details of the theoretical approach and a comparison with the electromagnetic (EM) inclusive responses of carbon 12 have been presented in~\cite{Franco-Munoz23}.
In this work, we extend that previous analysis by: i) studying the sensitivity of the calculations with the description of the knocked out nucleon, i.e., we address the effect of the treatment of final-state interactions and issues related with the (lack of) orthogonality between initial and final states; ii) comparing with inclusive cross sections; and iii) including calcium 40 in the analyses. 
Modeling and comparing with calcium 40 is interesting by itself, but also because it will pave the way to argon 40, which is the target material in a few neutrino detectors~\cite{DUNE16,MicroBooNE20}.

We find the contribution of MEC negligible in $R_L$ while it increases $R_T$ by around $30\%$. The increase in the cross section depends on the relative contribution of the transverse response, which depends on the kinematics.  
The agreement with $^{12}$C data is good in general and very good for $R_L$ for which isolation of the longitudinal contributions suppresses other processes not considered in the calculation, such as real pion production~\cite{Garcia-Marcos23}. Furthermore, the increase of the transverse component due to the two-body contribution is supported by the carbon data on the separated $R_T$ response.
Comparison of the predictions to data for separated responses for calcium, however, does not yield such clear conclusions. 
On the one hand, the agreement with the responses is globally poor. For the lowest available momentum transfer data (q=300, 350, 380  MeV/c), the predicted L response underestimates the data, while overestimating the T response. For the other kinematics, it is not so clear. 
Further, for the cross sections in calcium, the agreement is satisfactory, at the same level as that obtained for the carbon data.

In Sections~\ref{Sec:Oper} and \ref{Sec:Nuc-model}, we describe the interaction vertices and nuclear model.
In Sections~\ref{Sec:C-Resposes} and \ref{Sec:Ca-Resposes} we present our results for the inclusive EM responses from carbon and calcium. Cross sections are shown in Section~\ref{Sec:XS} and our conclusions are in~\ref{Sec:Conclu}.

\section{Theoretical model} \label{Sec:Theo}

The inclusive responses and cross sections are computed from the hadron tensor by integration over the variables of the final nucleon and sum over initial nucleons. 
The hadronic tensor is given by 
\be
    H_{\kappa}^{\mu\nu}= \sum_{m_j, s} [J_{\kappa,m_j,s}^\mu]^* J_{\kappa,m_j,s}^\nu,
\ee
where the hadronic current matrix element is 
\be
    J_{\kappa,m_j,s}^\mu = \int{d{\bf p}} \Psib^{s}({\bf p}+{\bf q}, \np_N) \Gamma^\mu \Psi_\kappa^{m_j}({\bf p}).
    \label{eq:hadroniccurrent}
\ee
$\np$ is the momentum of the bound nucleon, $\kappa$ represents the index of the nuclear initial state and $m_j$ the third-component of its total angular momentum $j$. $\np_N$ is the asymptotic momentum of the final nucleon knocked-out after the interaction with the boson, and $s$ its spin.

Bound nucleon momentum distribution $\Psi_\kappa^{m_j}$ components are obtained with the RMF model of~\cite{Sharma93}.  
For describing the final nucleon wave function $\Psi^s$, we use the energy-dependent relativistic mean-field (ED-RMF) potential, which is real, so that no flux is lost due to the imaginary part of the potential. The ED-RMF is the RMF potential used in the bound state but multiplied by a phenomenological function that weakens the potential for increasing nucleon momenta (see details in \cite{Gonzalez-Jimenez19,Nikolakopoulos19}). The main advantage of this choice is that it preserves the orthogonality between the initial and final states at low energies of the final nucleon, while approaching the behavior of the phenomenological optical potentials at larger energies. 

\begin{figure}[htbp]
\centering  
\includegraphics[width=0.4\textwidth,angle=0]{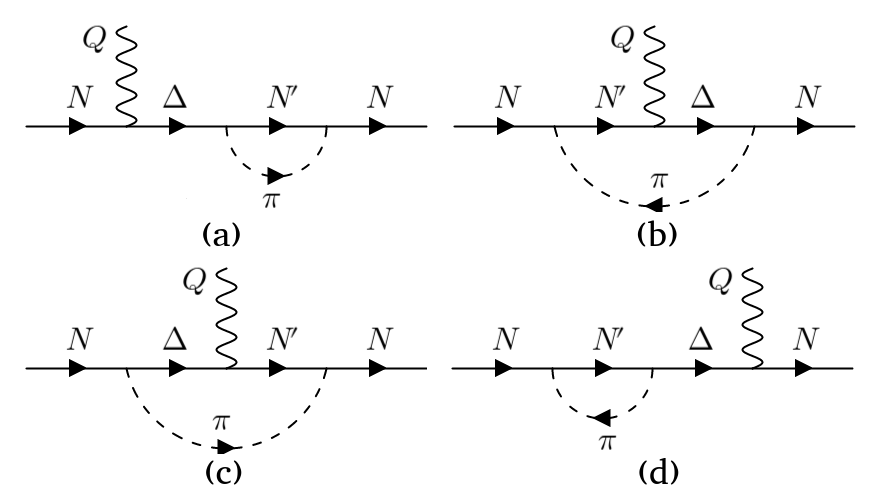}
\caption{Delta contributions. }
\label{fig:delta}
\end{figure}

\begin{figure}[htbp]
\centering  
\includegraphics[width=0.5\textwidth,angle=0]{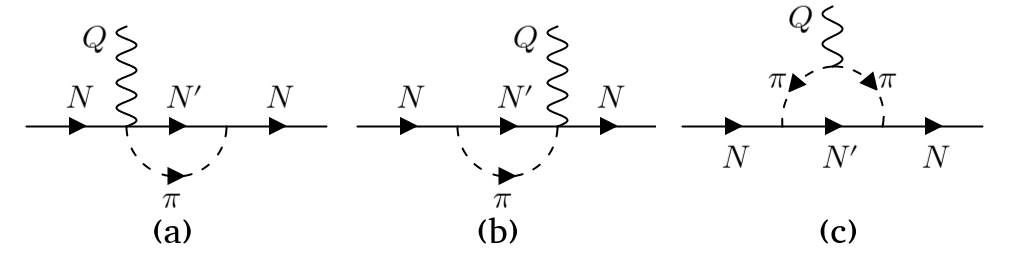}
\caption{Background contributions: seagull or contact [CT, (a) and (b)] and pion-in-flight [PF, (c)]. }
\label{fig:background}
\end{figure}

The matrix elements of the hadronic current operator $\Gamma^\mu$ of eq.~\ref{eq:hadroniccurrent} include all the processes that lead to a final 1p-1h state. 
Apart from the one-body current contributions, we include the contributions stemming from the two-body current operator that accounts for one-pion exchanged between interacting nucleons inside the nucleus, but only when the final state they produce correspond to 1p1h.
Thus, the hadronic matrix elements are coming from:
\ba
   \Gamma^\mu= \Gamma_{1b}^\mu + \Gamma_{2b}^\mu .
\ea
The one-body operator contribution is obtained from the usual CC2 prescription~\cite{Udias93,Udias95,Martinez06} for the current operator. 
The two-body contributions are the sum of the diagrams shown in Figs.~\ref{fig:delta} and \ref{fig:background}. 
They are discussed in what follows.

\subsection{Two-body current terms}\label{Sec:Oper}

We distinguish two different contributions to the two-body current: i) diagrams where a delta is involved, and ii) the background terms. When the 1p-1h excitation occurs through a two-body current, one of the outgoing nucleons of the two-particle two-hole interaction remains in a intermediate bound-nucleon state. Hence, the hadronic final state consists of just a nucleon. The complete 1p-1h contribution to the matrix element coming from the two-body operators is eventually obtained by integrating the intermediate state over all occupied levels in the initial nucleus. 
In the calculations presented here, the intermediate state is approximated by free Dirac spinors in a relativistic Fermi gas (RFG), as done in infinite nuclear matter~\cite{Amaro02}, but modified by an effective mass and energy that account for the relativistic interaction of nucleons with the mean-field potential~\cite{Franco-Munoz23}. 
The effective mass is given by
\be
M^*=m^* M <M,
\ee 
with $M$ the nucleon mass, and the modified nucleon energy is
\be
E^* = E + E_v,
\ee
where $E=\sqrt{p^2+(M^*)^2}$ is the on-shell energy with effective mass $M^*$. Following Ref. \cite{Martinez-Consentino21,Amaro18,Ivanov24}, we use an effective mass with $m^*=0.8$ and $0.73$ and a vector energy of $E_v=141$ and $210$ MeV for $^{12}$C and $^{40}$Ca, respectively.
We have verified that this simplified treatment of the intermediate state yields the same results as the calculation in which the intermediate states are described as discrete shell model bound states, but at a fraction of the computational cost~\cite{Franco-Munoz23}.

The two-body current is thus computed as 
\ba
  \Gamma_{2b}^\mu= \int{\frac{d\np_{ph}}{(2\pi)^{3}}} \Theta(p_F - p_{ph}) [\Gamma_{ChPT}^\mu + \Gamma_{\Delta}^\mu ],
\ea
where $\Gamma_{\Delta}^\mu$ and $\Gamma_{ChPT}^\mu$ are the contributions from the diagrams in Figs. \ref{fig:delta} and \ref{fig:background}, respectively. The explicit expressions of the operators and further details can be found in \cite{Franco-Munoz23}.

\subsection{Nuclear model}\label{Sec:Nuc-model}

In an independent-particle shell model (IPSM), $^{12}$C is made of 2 and 4 nucleons in the $1s_{1/2}$ and $1p_{3/2}$ states and accordingly for $^{40}$Ca, as shown in Table \ref{tab:occupations-40Ca}.
Each shell has a unique binding energy, which means that the missing energy distribution~\footnote{By missing energy we refer to the part of transferred energy that transforms into internal energy of the residual nucleus.} predicted by the model would be the sum of one Dirac delta per shell, normalized to its occupancy. 
However, the extreme shell model is a very crude approximation to the missing energy distribution of the strength, which has been measured in $(e,e'p)$  and $(p,2p)$ experiments for carbon, calcium and other nuclei~\cite{Dutta03, Fissum04, Kramer89, Yasuda10, Volkov90, Jiang22}. It is also observed that the occupancy of the shells is depleted with respect to the independent-particle shell-model predictions~\cite{Udias93, Giusti11, Atkinson18}, and that these `missing nucleons' re-appear in deeper missing-energy ($E_m$) and missing-momentum ($p_m$) regions~\cite{Egiyan06,Duer18}.
This behavior is due to effects beyond the independent-particle approach, ascribed to short- and long-range correlations~\cite{Mahaux87,Ma91,DelAtti15,Colle16}. 
To say it in a few words, whenever there is enough energy transferred to the nucleus, so that more than one nucleon may be knocked out (that is, beyond the two nucleon emission threshold), correlations in the initial state would make possible the emission of additional nucleons (spectators in the impulse approximation), which will carry  energy and momentum. The analysis of $(e,e'p)$, both below and above the two nucleon threshold, made it possible to obtain a semi-phenomenological spectral function~\cite{Benhar94,Ankowski24}, composed from discrete and continuum contributions. The discrete spectral function arises from the mean field contribution, partly depleted by correlations, which shift strength (that is, probability of knocking out nucleons from the initial state) to higher excitation energies, among other reasons because correlations allow to knock-out additional nucleons, even if the exchanged boson interacts only with one nucleon. As introduced in reference~\cite{Gonzalez-Jimenez22}, we take the semi-phenomenological spectral function, in the case of carbon, the Rome spectral function~\cite{Benhar94,Benhar05}, and substitute its discrete component by the one given by RMF single-particle states. We take occupations as in~\cite{Franco-Patino22}, and we employ a continuous missing-energy profile. The occupation of the shells is 3.3 ($1p_{3/2}$) and 1.8 ($1s_{1/2}$) nucleons.

For $^{40}$Ca, the missing-energy profile has been constructed taking as reference the results from several analysis \cite{Kramer89,Volkov90,Yasuda10,Atkinson18,Udias93,Giusti11,Jiang22}. 
For the most external shell, $1d_{3/2}$, we use the separation energy from nuclear mass tables~\cite{NuclearMasses} to set the peak in the proton and neutron distributions. For the internal shells, the peaks of the proton distribution have been taken from \cite{Volkov90},  and for neutrons we shift the proton values to account for the Coulomb interaction. 
This shift is calculated as the difference between the neutron and proton separation energies. 
As we made for carbon~\cite{Franco-Patino22}, we model the peaks as Gaussian distributions, narrower for the external shells, and wider for the deeper ones. We stress that the inclusive observables presented in this work, responses and cross sections, are mostly sensitive to the overall occupation probability of the shells and not to fine details of the missing-energy profile.
Finally, in calcium, we have introduced occupation ranges to account for the uncertainties and discrepancies of the different studies, the used values are shown in Table~\ref{tab:occupations-40Ca}. We use the same occupations for protons and neutrons in both carbon and calcium cases. 

\begin{table}[ht]
    \centering
    \begin{tabular}{|c|c|c|c|c|c|c|c|}
        \hline
         &  & \multicolumn{6}{c|}{$n_\alpha$}\\
        \cline{3-8}
        $\alpha$ & $N_\alpha$ & (e,e'p)  & (e,e'p) & (e,e'p) & (e,e'p)& (p,2p) & This \\
         & & \cite{Kramer89} &  \cite{Atkinson18} &  \cite{Udias93} & \cite{Giusti11} & \cite{Yasuda10} & work \\
        \hline
        $2s_{1/2}$ & 2 & 0.58 - 0.7  & 0.57 - 0.63  &  0.48 - 0.54 &  0.51        & 0.49 - 0.57 & 0.5 - 0.7 \\
        $1d_{3/2}$ & 4 & 0.58 - 0.72 & 0.67 - 0.75  &  0.72 - 0.8  &  0.49 - 0.69 & 0.6 - 0.7   & 0.5 - 0.7 \\
        $1d_{5/2}$ & 6 & 0.78 - 0.88 &              &              &              & 0.76 - 0.94 & 0.6 - 0.8 \\
        $1p_{1/2}$ & 2 &             &              &              &              & 0.42 - 0.56 & 0.6 - 0.8 \\
        $1p_{3/2}$ & 4 &             &              &              &              & 0.42 - 0.56 & 0.6 - 0.8 \\
        $1s_{1/2}$ & 2 &             &              &              &              & 0.8 - 0.98  & 0.7 - 0.85 \\
        \hline
    \end{tabular}
    \caption{$^{40}$Ca shell model states ($\alpha$), their occupations according to the independent-particle shell model ($N_\alpha$)  and occupation probabilities ($n_\alpha$) from the references \cite{Kramer89,Atkinson18,Udias93,Giusti11,Yasuda10} and the ranges used in this work. }
    \label{tab:occupations-40Ca}
\end{table}

Furthermore, the high missing energy and momentum region of the spectral function due to short-range correlations (SRC), representing the continuum contribution to the spectral function, is also included. The momentum distribution is modeled as a $s$-wave fitted to reproduce the momentum distribution of the Rome spectral function and normalized so that after summing over all shells the 12 and 40 nucleons of $^{12}$C and $^{40}$Ca, respectively, are recovered. 
Its contribution starts at the two-nucleon emission threshold, it has a soft maximum at around 100 MeV and an exponential fall that extend to high momentum~\cite{Gonzalez-Jimenez22,Franco-Patino22}. 
In what follows, we refer to this contribution as SRC background. The contribution of this SRC term to the inclusive cross-sections or responses for the kinematical settings included in this work, is anyway very insensitive to the actual shape of its momentum distribution. 

It is worth noticing that in the case that we introduce this representation of the spectral functions in our formalism, without two body matrix elements and  in PWIA (that is, in a factorized calculation), the standard result of the spectral function is essentially recovered \cite{Gonzalez-Jimenez22}.

\section{Results}\label{Sec:Res}

\subsection{Carbon responses}\label{Sec:C-Resposes}
\begin{figure}[htbp]
\centering  
\includegraphics[width=0.48\textwidth,angle=0]{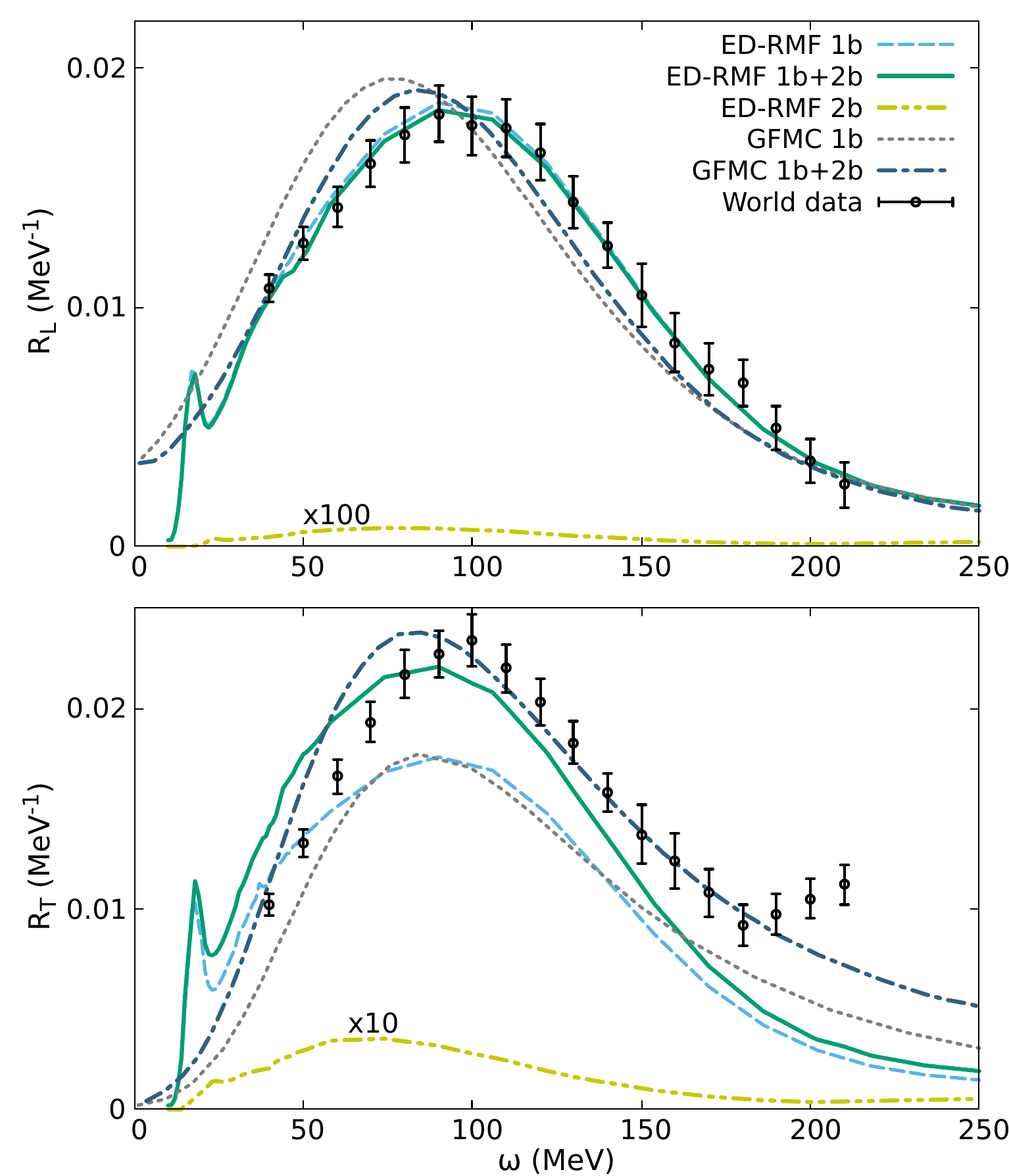}
\caption{$^{12}$C longitudinal (up) and transverse (bottom) electromagnetic inclusive response functions. The transferred momentum $q$ is 380 MeV/$c$. We show our ED-RMF results and the GFMC responses taken from~\cite{Lovato16}. The longitudinal and transverse two-body contributions are also shown and multiplied by a factor 100 and 10, respectively.}
\label{fig:RL-RT}
\end{figure}

In Fig.~\ref{fig:RL-RT} we show our results for the inclusive longitudinal and transverse responses of $^{12}$C, computed with one- and two-body operators. Theoretical responses are compared to experimental data extracted by means of a Rosenbluth analysis by Jourdan \cite{Jourdan96} and Barreau et al. \cite{Barreau83}.  We also show the {\it ab initio} non-relativistic Green's function Monte Carlo (GFMC) responses of~\cite{Lovato16}. 
We highlight the following features. The main effect of the two-body currents with respect to the one-body approach appears in the transverse channel, while in the longitudinal one the 1p-1h MEC contribution is hardly visible. The transverse response increases by up to 30\% for ED-RMF.
The agreement of our results with data is good, outstanding for the longitudinal response. It is also remarkable the good agreement between ED-RMF and GFMC calculations, despite the fact that they represent quite different theoretical approaches. 

Additionally, this same Fig.~\ref{fig:RL-RT} also shows the two-body contribution only. For the transverse and longitudinal responses, it is nearly two and four orders of magnitude smaller than the one-body contribution, respectively. Therefore, it is clear that the increase of the transverse response is due to the interference between the one- and two-body contributions.

\begin{figure}[ht!]
\centering  
\includegraphics[width=0.49\textwidth,angle=0]{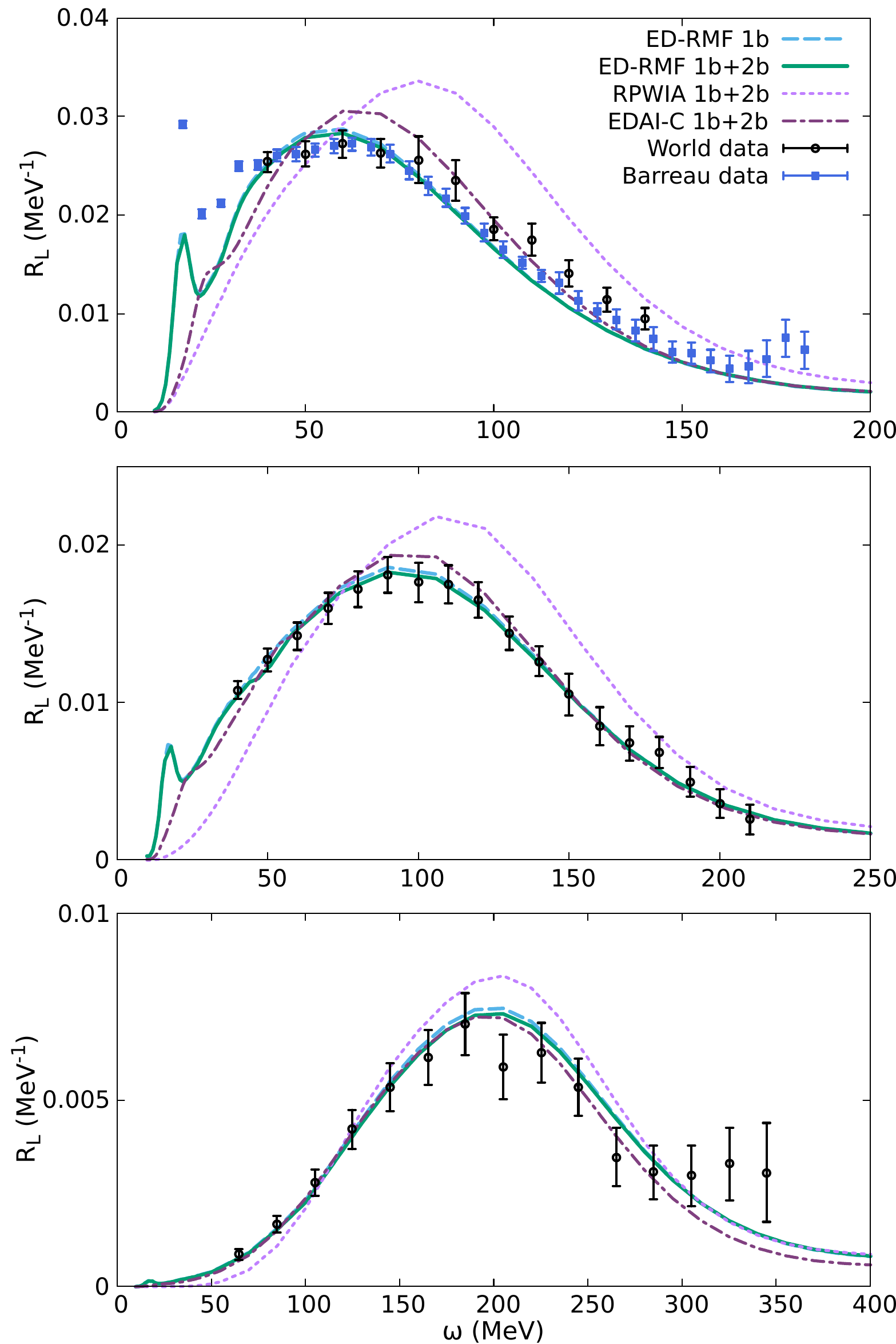}
\includegraphics[width=0.49\textwidth,angle=0]{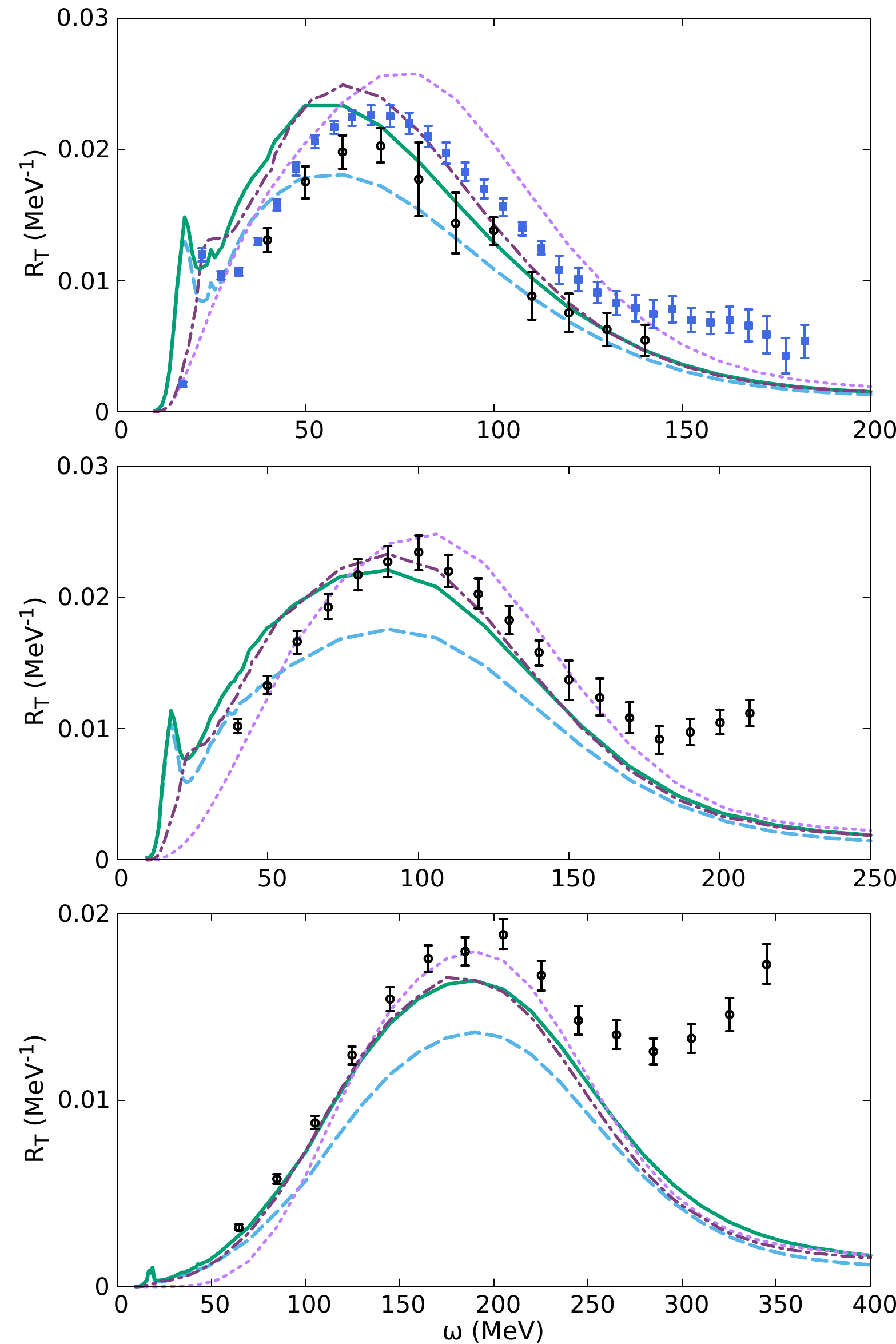}
\caption{L (left) and T (right) responses of $^{12}$C with the ED-RMF, RPWIA and EDAI-C models. The transferred momentum $q$ is (from up to bottom) 300, 380 and 570 MeV/$c$.}
\label{fig:RL-RT-potentials}
\end{figure}

In Fig.~\ref{fig:RL-RT-potentials} we study the distortion of the knocked out nucleon and the spurious contributions arising from the non-orthogonality between initial and final states~\cite{Gonzalez-Jimenez19} in carbon.
We show results with the so-called relativistic plane wave impulse approximation (RPWIA) model, in which the final nucleon is described by a relativistic plane wave. As expected, RPWIA calculations overestimate the data, more so at small values of transfer momentum. Further, the low energy behavior of the responses is very different from the data.  This may be attributed to the distortion of the final nucleon that is not being included in RPWIA, which, among other effects, implies a lack of orthogonality between initial and final states, giving rise to spurious contributions to the responses.
The effect of distortion, as shown in the ED-RMF approach is to shift the peak of the responses to the correct position, according to the data, to reduce the total strength and to redistribute it from the peak to the tails, particularly to the low energy end of the responses, causing the ED-RMF to reproduce the data remarkably well.  

Further, we compare ED-RMF results with those obtained treating the distortion of the final nucleon with the real part of the phenomenological  energy-dependent A-independent carbon relativistic optical potential EDAI-C~\cite{Cooper93}. 
This potential was extracted by fitting elastic proton-carbon scattering data in the range $30<T_p<1040$ MeV, $T_p$ being the proton kinetic energy. 
The two approaches (ED-RMF and EDAI-C) provide very similar results for large enough values of the momentum transfer, $q>300$ MeV$/c$~\cite{Gonzalez-Jimenez20}. However, the EDAI-C, unlike the ED-RMF, does not preserve exact orthogonality between the initial and final states; hence, when the momentum of the final nucleon is comparable to the momentum of the bound nucleon (i.e., approximately up to $p_N<300$ MeV$/c$), the spurious non-orthogonality contributions become an issue for EDAI-C as well as for RPWIA. 
This is confirmed by our results, in which one observes that even though EDAI-C and ED-RMF are similar both in shape and magnitude, the agreement with the data is slightly better for ED-RMF, specially, at the low energy tails. Taking this into account, the remaining results from this work have been computed with the ED-RMF potential.

\subsection{Calcium responses}\label{Sec:Ca-Resposes}

\begin{figure}[ht!]
\centering  
\includegraphics[width=0.9\textwidth,angle=0]{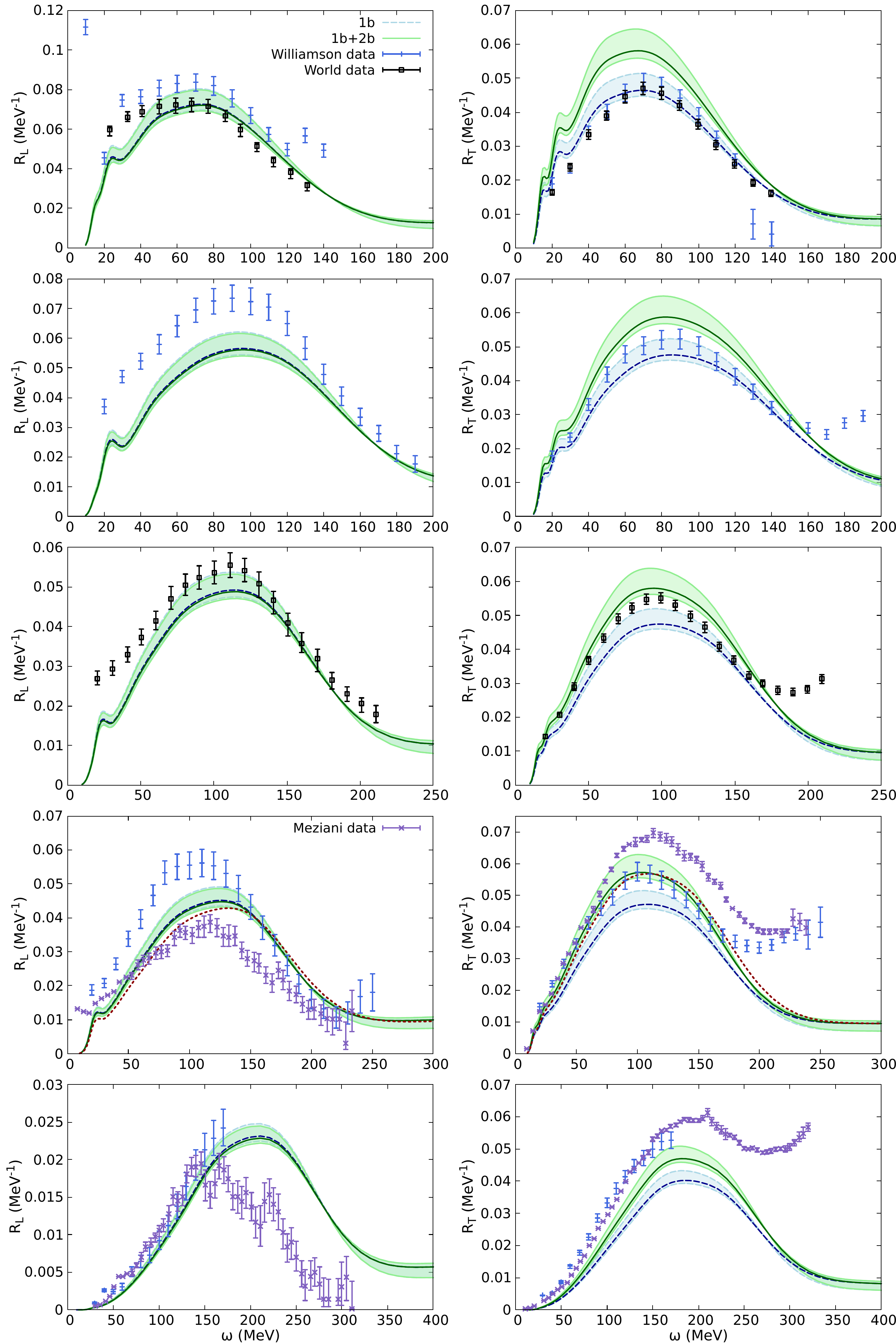}
\caption{$^{40}$Ca longitudinal (left) and transverse (right) electromagnetic inclusive responses using the ED-RMF model. The transferred momentum $q$ is (from up to bottom) 300, 350, 380, 400 and 550 MeV/$c$. In the case of $q=400$ MeV/$c$, Meziani data and red line correspond to $q=410$ MeV/$c$, the theory line was computed with 1b+2b currents.} 
\label{fig:RL-RT-calcio}
\end{figure}

The inclusive responses of $^{40}$Ca are shown in Fig.\ref{fig:RL-RT-calcio}. 
Our results are now presented as bands that account for the uncertainty in the occupation probability of the single-particle states. 
We have generated $10^4$ random combinations for the $^{40}$Ca occupations within the limits shown in Table \ref{tab:occupations-40Ca}, with the constraint that the amount of nucleons in the SRC background is between $20\%$ and $35\%$ of the total number of nucleons~\footnote{This strategy is similar to the one employed in~\cite{Franco-Patino24} to model the uncertainties in the description of the initial-state spectral function of argon 40.}
The two extremes of these $10^4$ combinations, which correspond to the limits of the bands, are those with the most and least nucleons in the SRC background within the allowed values. 
An additional dark-solid line is shown, it corresponds to the case with the mean value of the occupations of Table~\ref{tab:occupations-40Ca}, this means 32\% of the nucleons in the SRC background.
The results are compared to the available experimental data from \cite{ Jourdan96,Meziani84, Meziani85,Williamson97}. 

For the longitudinal responses, at momentum transfer $q=300$ MeV/c, the data from Jourdan \cite{Jourdan96} and Williamson \cite{Williamson97} agree fairly well with each other, and with our results. 
At $q=350$ MeV/c we underestimate the quasielastic peak with respect to the predictions from Williamson \cite{Williamson97}. 
Meanwhile, at $q=380$ MeV/c, there is a good agreement with Jourdan data \cite{Jourdan96}. 
Regarding the transverse response, for these three kinematics, we observe a good agreement of our one-body results with the experimental data, therefore, when we introduce the two-body currents the data is clearly overestimated. 
Our one-body current results in the transverse sector are similar to those obtained with the mean-field approach of \cite{Jachowicz21} and 
the different \textit{ab initio} approach of \cite{Sobczyk21,Sobczyk24}. 
Hence, it is expected that these other calculations show an overestimation of the responses similar to what we get here when the two-body currents are included.
We stress that the transverse enhancement due to two-body currents is supported by the analyses of inclusive EM responses for $^3$He, $^4$He and $^{12}$C~\cite{Carlson02,Lovato15,Andreoli22,Lovato16,Franco-Munoz23}.

The fourth kinematic corresponds to $q=400$ MeV/c, which is compared with the data at $q=400$ MeV/c from \cite{Williamson97} and $q=410$ MeV/c from \cite{Meziani84,Meziani85}. Our longitudinal result lies in between both data sets, which are in large disagreement, meanwhile the transverse response including two-body currents seems to agree better with \cite{Meziani85}, although it would be necessary to have a better control of the occupations and then obtain a narrower band to draw a clear conclusion. 
The large difference between the two data sets is remarkable and cannot be explained by the 10 MeV difference in momentum transfers, which is illustrated by showing in the same figure our 1b+2b current calculation for $q=410$ MeV/c (short-dashed red line, the mean value of the occupations was employed).

For the last momentum transfer considered, $q=550$ MeV/c, there is a fair agreement between our results and the data from \cite{Williamson97} and \cite{Meziani84} up to an energy transfer around 150 MeV. 
In the transverse sector, the situation is similar to the results obtained in the case of carbon, with an increase of the transverse response up to 25$\%$ which improves the agreement with data. 

We recall that the extraction of the responses from the inclusive cross section data, has been done with a Rosenbluth procedure, which is very sensitive to the treatment of Coulomb distortion effects in the electrons.
In carbon this is expected to be a very small correction. But in calcium it is a non-negligible effect which introduces uncertainties that are difficult to quantify. The disagreement among different data sets for the separated responses in calcium may point to unadequate treatment of Coulumb distortion. For this reason, in the next section we compare directly with inclusive cross section data.

\subsection{Cross sections}\label{Sec:XS}

\begin{figure}[htbp]
\centering  
\includegraphics[width=\textwidth,angle=0]{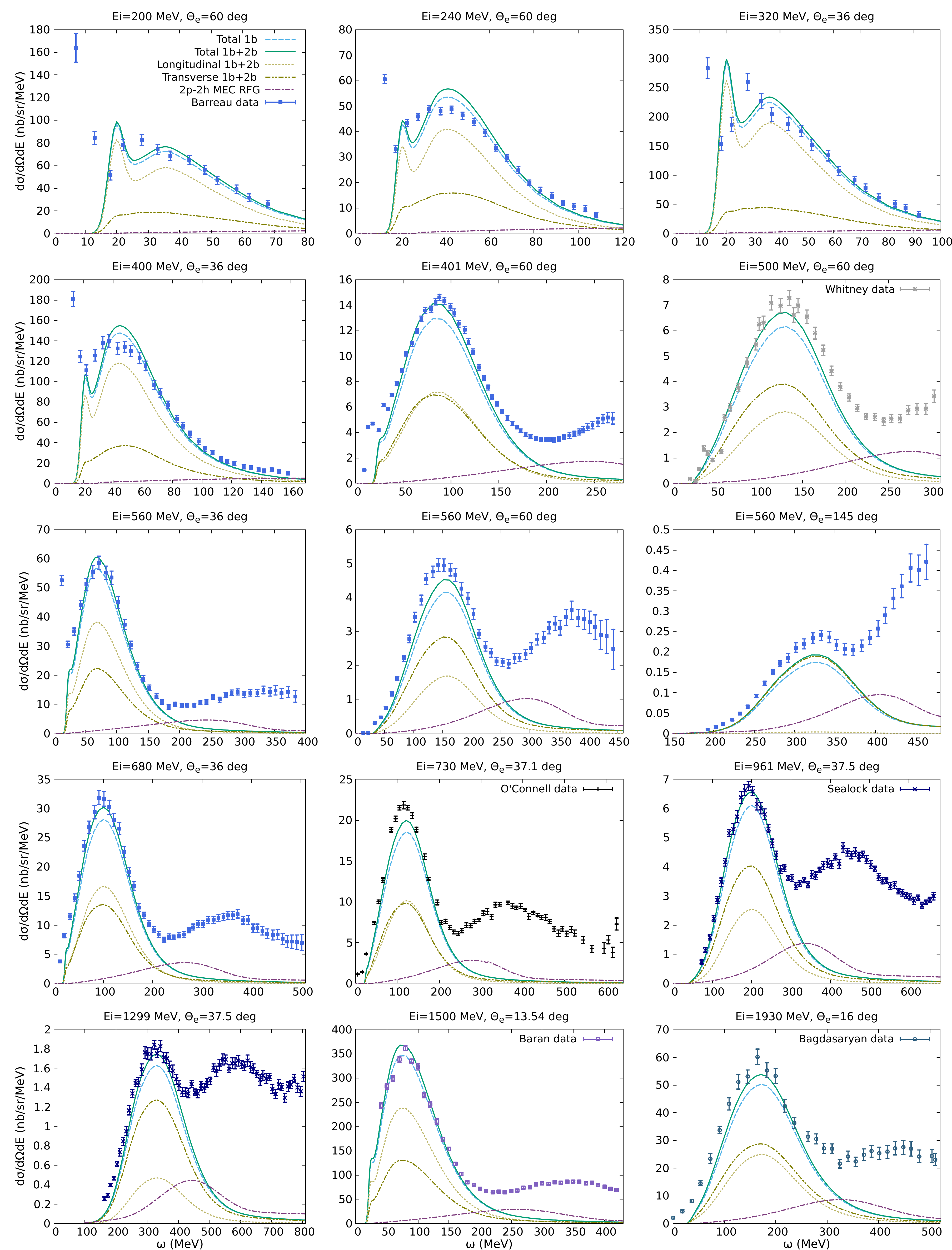}
\caption{$^{12}$C electromagnetic inclusive cross sections at various beam energies and scattering angles. We show the 1b and 1b+2b ED-RMF cross section (total), its longitudinal and transverse contributions, and the MEC 2p-2h from~\cite{Megias15}.}
\label{fig:cross-section}
\end{figure}

\begin{figure}[htbp]
\centering  
\includegraphics[width=\textwidth,angle=0]{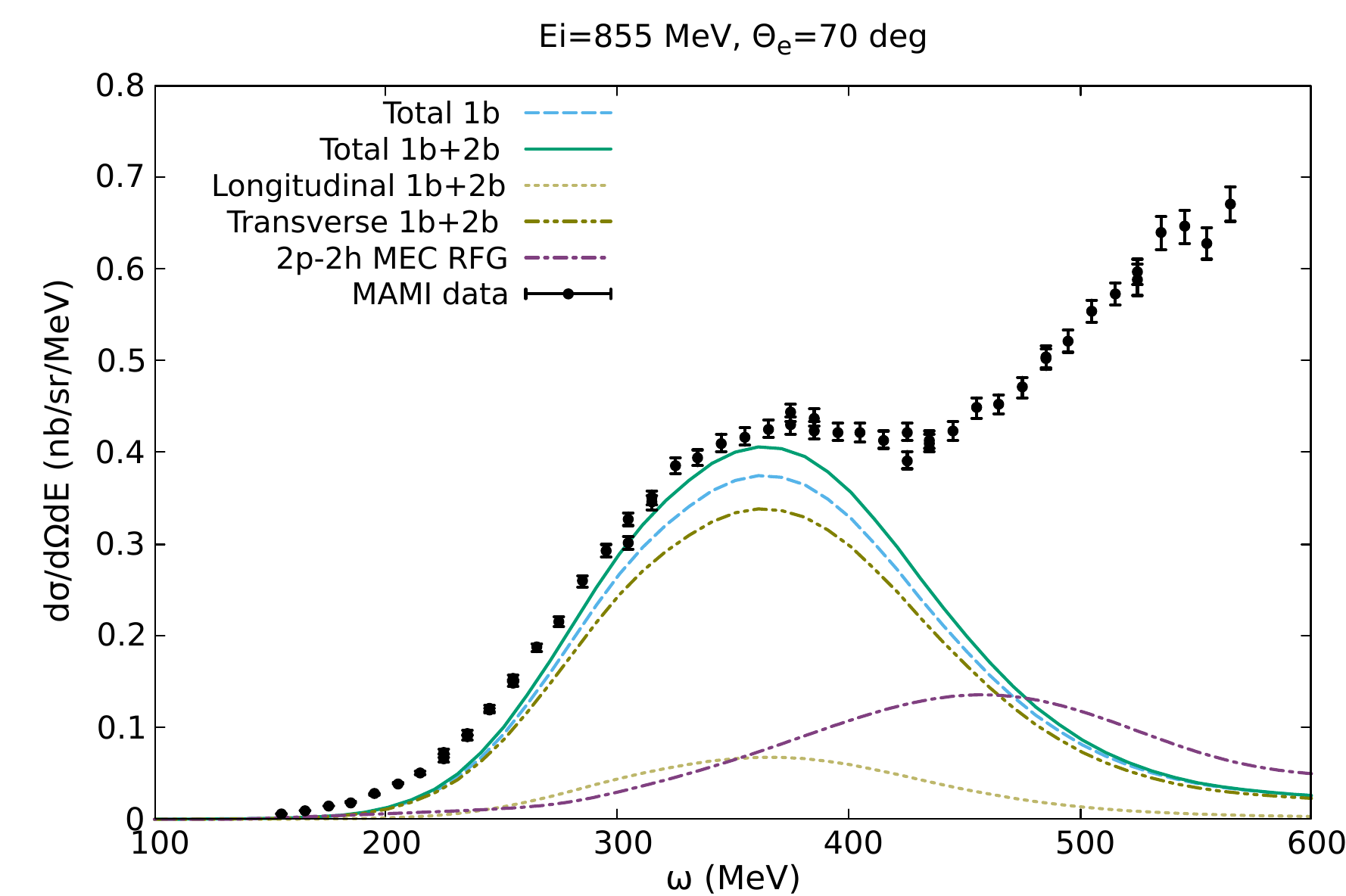}
\caption{$^{12}$C electromagnetic inclusive cross sections at beam energy 855 MeV and scattering angle 70 deg. We show the 1b and 1b+2b ED-RMF cross section (total), its longitudinal and transverse contributions, and the MEC 2p-2h from~\cite{Megias15}.}
\label{fig:cross-section-MAMI}
\end{figure}

\begin{figure}[htbp]
\centering  
\includegraphics[width=\textwidth,angle=0]{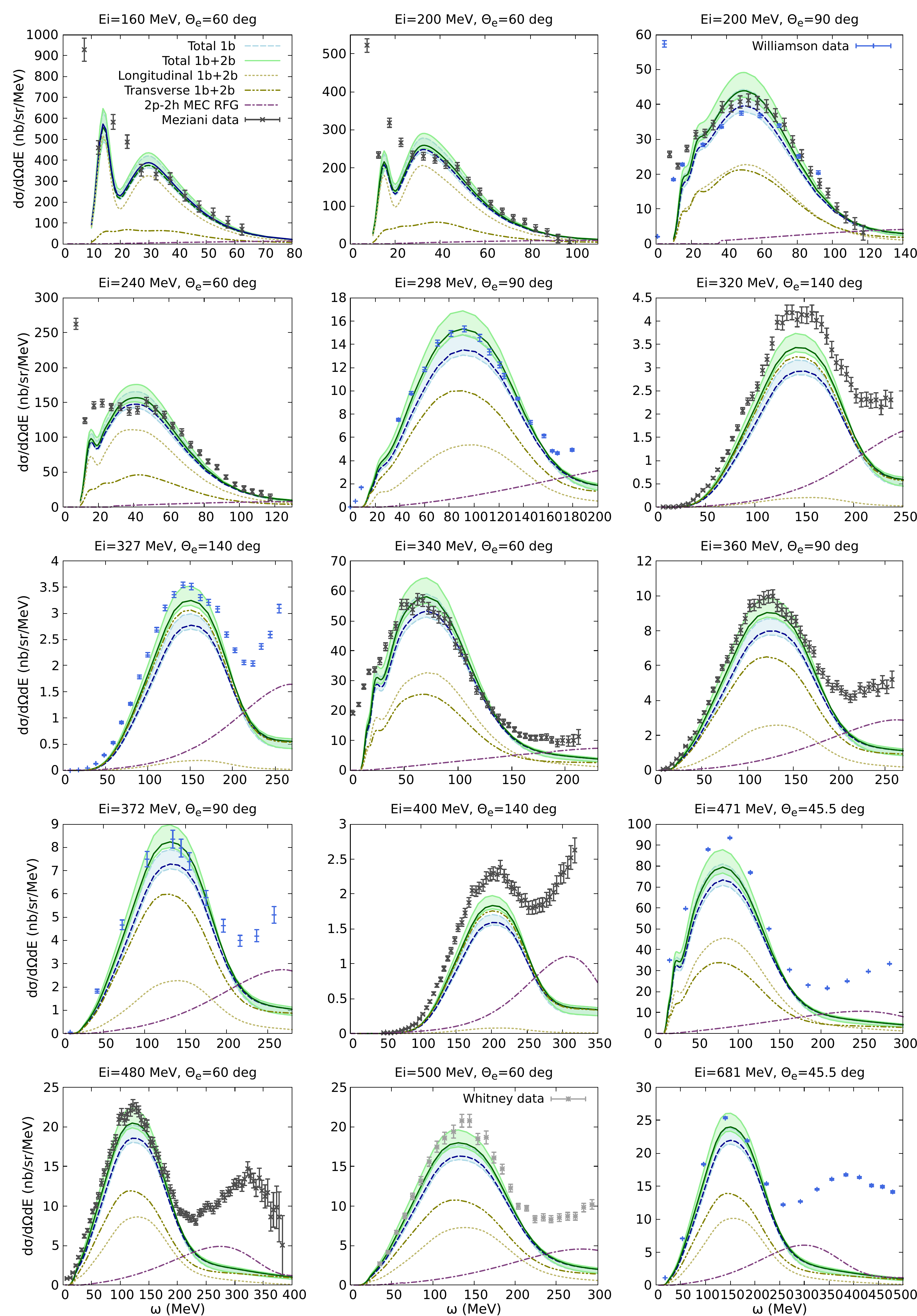}
\caption{$^{40}$Ca electromagnetic inclusive cross sections at various beam energies and scattering angles. We show the 1b and 1b+2b ED-RMF cross section (total), its longitudinal and transverse contributions, and the MEC 2p-2h from~\cite{Megias15}. The bands represent the uncertainty in the occupation of the single-particle states according to Table~\ref{tab:occupations-40Ca}.}
\label{fig:cross-section-40Ca}
\end{figure}

In Figs.~\ref{fig:cross-section}, \ref{fig:cross-section-MAMI} and \ref{fig:cross-section-40Ca} we show inclusive electron-$^{12}$C and -$^{40}$Ca cross sections at various beam energies and scattering angles. 
Our QE 1b and 1b+2b ED-RMF predictions are compared to experimental data from \cite{Barreau83,Whitney74,O'Connell87,Sealock89,Baran88,Bagdasaryan88,Mihovilovic24} for carbon and \cite{Meziani84,Williamson97,Whitney74} for calcium. 

For the studied kinematics, the two-body contributions to the 1p1h produce an increase of the cross section coming from the increase in the transverse part.  The relative increase is larger for larger values of beam energy, $E_i$, and lepton scattering angle, $\theta_e$, as the contribution of the transverse response is more significant in those cases. 
We stress that our calculation is only for QE 1p-1h channel, Thus there is strength in the experimental data unaccounted for in our prediction, coming from other reaction channels, such as  2p-2h contributions to the final state, and pion production (see e.g. the review article~\cite{Amaro20}). 
Aiming at making the comparison more meaningful we included the MEC 2p-2h contribution (dashed-dotted purple line) from~\cite{DePace03,Megias15}, which should be added incoherently to the QE contribution. 
We point out that this MEC 2p-2h corresponds to a relativistic Fermi gas model, it is expected that a shell model calculation of this channel yields a somewhat smaller contribution~\cite{VanCuyck17,Niewczas-PhD}.

Since the longitudinal part of the cross section is basically not affected by the two-body currents, 
we identify kinematics with a small transverse contribution to benchmark our  occupation probabilities.
For carbon, Fig.~\ref{fig:cross-section}, the panels corresponding to  $E_i=200$ MeV at $\theta_e=60$ deg, and $320$ and $400$ MeV at $36$ deg are mostly longitudinal. In these cases we find good agreement with the data. 
For calcium, Fig.~\ref{fig:cross-section-40Ca}, the most longitudinal kinematics correspond to $E_i=160$, $200$ and $240$ MeV at $60$ deg. 
The agreement with data is good for energies larger than 30 MeV, where the quasielastic contribution dominates. For  energies below 30 MeV, there is qualitative agreement, but in this region one would have to include discrete resonances in the calculation to agree with data.

Analogously, we identify kinematics that are essentially transverse. For carbon, these are $560$ MeV at 145 deg and the recent measurement from the Mainz Microtron (MAMI) experiment at 855 MeV and 70 deg \cite{Mihovilovic24} shown in Fig.\ref{fig:cross-section-MAMI}.  
For calcium, these are $320$, $327$ and $400$ MeV at 140 deg. 
In all these cases, the data support the enhancement of the cross section due to the 2-body contributions.

In the case of calcium, for most of the kinematics studied, the upper part of our 1b+2b bands is often above the data or it would be if one introduces the contribution from other reaction channels, such as MEC 2p-2h.  
In other words, our results fit better the data when around $30-35\%$ of the nucleons are placed in the SRC background that contributes mostly in the high-$\omega$ tail and little in the QE peak.
Recent works have estimated about $20\%$~\cite{Egiyan06,Duer18} of SRC pairs in nuclei, therefore,
the relatively small occupation probabilities that our comparison with data suggests and, consequently, our large SRC background are effectively accounting for other mechanisms, not included in our approach, which lead to a reduction of the cross section in the QE peak. 

Finally, we discuss different $^{40}$Ca data sets corresponding to the same lepton scattering angle and similar or identical beam energy, aiming at assessing possible inconsistencies between them.
At 140 deg and 320 and 327 MeV, 90 deg and 200 MeV, and 90 deg and 360 and 372 MeV we found data sets from Meziani~\cite{Meziani84} (black points) and Williamson~\cite{Williamson97} (blue points). 
In these cases, we obtain a better agreement with Meziani's sets, which are systematically larger than the ones from Williamson. 
At 60 deg and 480 and 500 MeV we found data sets from Meziani and Whitney~\cite{Whitney74} (gray points), which seem to be consistent with each other and with our predictions, with the intermediate value for the occupations.

\section{Conclusions}\label{Sec:Conclu}

Our relativistic mean-field based model, with one- and two-body current contributions to the 1p-1h QE peak, can simultaneously describe the inclusive longitudinal and transverse electromagnetic responses of $^{12}$C in the quasielastic regime, and further in a good agreement with the experimental cross sections. 

The results for $^{40}$Ca are less conclusive. The agreement of our model with $R_L$ and $R_T$ data is not good, but there are clear discrepancies between different data sets, both in the cross sections and, specially, in the responses.
The treatment of the interaction of the electron with the Coulomb field affects the Rosenbluth separation, causing a redistribution of the strength attributed to longitudinal and transverse responses and changing their shape~\cite{Giusti87,Jin92,Udias93}. This leads to significant uncertainties in the extracted responses. 
Therefore, the study of the inclusive cross sections seemed to be more convenient in this case.  Within the margins we chose for the shell occupations, we may overestimate or agree with the data. 
For most of the kinematics studied the data favour smaller shell occupations, with around $30-35\%$ of the nucleons in the SRC background. 
This makes clear the relevance of short-range correlations, as well as other mechanisms not included in our approach, such as long-range correlations, that reduce the strength of QE peak.

The key contribution of this work is the incorporation of the two-body meson-exchange current contribution to the 1p-1h channel. It includes the delta resonance mechanism and background terms. 
We find that the effect of the two-body currents is only significant in the transverse channel, where the response is increased up to a 30\%. The delta resonance mechanism is the main responsible of this result, giving the larger contribution. 

This work paves the way for the leap to neutrino-nucleus interaction processes and to modeling of lepton-$^{40}$Ar scattering. We point out that in the case of charge-current quasielastic (anti)neutrino reactions the transverse response is clearly the dominant one~\cite{Megias14,Megias15}, except at very low four-momentum transfer. Therefore, we expect the two-body  contributions to play an important role in the neutrino sector.

\section*{Acknowledgements}
This work was supported by the Madrid Government under the Multiannual Agreement with Complutense University in the line Program to Stimulate Research for Young Doctors in the context of the V PRICIT (Regional Programme of Research and Technological Innovation), project PR65/19-22430; 
by project PID2021-127098NA-I00 funded by MCIN/AEI/10.13039/501100011033/FEDER, UE;
and 
by project RTI2018-098868-B-I00 (MCIN/AEI,FEDER,EU). 
The computations of this work were performed in Brigit, the HPC server of the Complutense University of Madrid.

\section*{References}
\bibliography{bibliography}
\bibliographystyle{unsrt.bst}

\end{document}